\documentclass[traditabstract]{aa} 
\usepackage{graphicx}
\usepackage{txfonts}
\usepackage{natbib}
\usepackage{soul}
\bibpunct{(}{)}{;}{a}{}{,} 
\usepackage[normalem]{ulem}

\begin{document}
   \title{A short-period super-Earth orbiting the M2.5 dwarf GJ\,3634 \thanks{Based on observations made with the \textsc{Harps} instrument on the ESO 3.6-m telescope at La Silla Observatory under program IDs 082.C-0718(B) and183.C-0437(A), and observations made with {\it Warm Spitzer} under program 60027. Radial-velocity and photometric tables are only available in electronic form at the CDS via anonymous ftp to cdsarc.u-strasbg.fr (130.79.128.5) or via http://cdsweb.u-strasbg.fr/cgi-bin/qcat?J/A+A/}}
   \subtitle{Detection with {\it \textsc{Harps}} velocimetry and transit search with {\it Spitzer} photometry}

	\author{X. Bonfils \inst{1,2}
		\and M. Gillon \inst{2,3}
	         \and T. Forveille \inst{1}
	         \and X. Delfosse \inst{1}
	         \and D. Deming \inst{4}
	         \and B.-O. Demory \inst{5}
	         \and C. Lovis \inst{2}
	         \and M. Mayor \inst{2}
	         \and V. Neves \inst{1,6}
	         \and C. Perrier \inst{1}
	         \and N. C. Santos \inst{6,7}
	         \and S. Seager \inst{5}
	         \and S. Udry \inst{2}
	        \and I. Boisse \inst{6}
	        \and M. Bonnefoy \inst{1}
		}

	\offprints{X. Bonfils\\ \email{Xavier.Bonfils@obs.ujf-grenoble.fr}}
	
	\institute{UJF-Grenoble 1 / CNRS-INSU, Institut de Plan\'etologie et d'Astrophysique de Grenoble (IPAG) UMR 5274, Grenoble, F-38041, France.
          \and Observatoire de Gen\`eve, Universit\'e de Gen\`eve, 51 ch. des Maillettes, 1290 Sauverny, Switzerland
		\and Institut dÕAstrophysique et de G\'eophysique, Universit\'e de Li\`ege, All\'ee du 6 Ao\^{u}t 17, Bat. B5C, 4000 Li\`ege, Belgium
 \and Planetary Systems Branch, Code 693, NASA/Goddard Space Flight Center Greenbelt, MD 20771, USA
\and Department of Earth, Atmospheric and Planetary Sciences, Department of Physics, Massachusetts Institute of Technology, 77 Massachusetts Ave., Cambridge, MA 02139, USA
       \and
        Centro de Astrof{\'\i}sica, Universidade do Porto, Rua das Estrelas,
	  P4150-762 Porto, Portugal
	  \and
	  Departamento de F\'isica e Astronomia, Faculdade de Ci\^encias, Universidade do Porto, Portugal
		}

	\date{Received/Accepted}
	
	\abstract
	{We report on the detection of GJ\,3634b, a super-Earth of mass $m \sin i = 7.0_{-0.8}^{+0.9}~\mathrm{M_\oplus}$  and period $P = 2.64561 \pm 0.00066 $~day. Its host star is a M2.5 dwarf, has a mass of $0.45\pm0.05~\mathrm{M_\odot}$, a radius of 0.43$\pm$0.03 $\mathrm{R_\odot}$ and lies $19.8\pm0.6$ pc away from our Sun. The planet is detected after a radial-velocity campaign using the ESO/\textsc{Harps} spectrograph. GJ\,3634b had an {\it a priori} geometric probability to undergo transit of $\sim$7\% and, if telluric in composition, a non-grazing transit would produce a photometric dip of $\lesssim$0.1\%.  We therefore followed-up upon the RV detection with photometric observations using the 4.5-$\mu$m band of the IRAC imager onboard {\it Spitzer}. Our six-hour long light curve excludes that a transit occurs for 2$\sigma$ of the probable transit window, decreasing the probability that GJ\,3634b undergoes transit to $\sim$0.5\%.}


   \maketitle
%

\section{Introduction}
The subset of extrasolar planets that transit their parent star have had the most impact on our understanding of their planetary structure and atmospheric physics \citep[as reviewed by][]{Charbonneau:2007}. They are the only ones for which one can simultaneously measure mass and radius, and, by inference, internal composition. The few that transit a host star bright enough for detailed  spectroscopic follow-up provided additional observational information on the composition and physics of extrasolar planetary atmospheres, which opened the new scientific field of physical exoplanetology. That select group of very bright transiting systems, with transit depths deep enough for detailed characterization, only has a handful of members, and until recently all were gaseous giant planets. The recent discoveries that GJ\,436b and GJ\,1214b undergo transits \citep{Gillon:2007b, Charbonneau:2009} has extended that new field to the realms of the ice giants and super-Earth planets. 

To search for transiting planets, two strategies compete: the {\it photometric} and the {\it radial-velocity educated} approach. The photometric approach detects planets when they transit their parent star and, {\it de facto}, is the most direct strategy to find transiting planets. Alternatively however, one may wait for the detection of a planet before undertaking its photometric search for transit. If the planet is first detected, with the radial-velocity (RV) technique for instance, not only is the presence of the planet then known for sure, but the observational window to perform a photometric search is very much narrowed with an a priori ephemeris. This latter approach, though less direct, can prove more successful in finding planets that transit bright nearby stars, like for the ``blockbusters''  HD209458b, HD189733b and GJ436b, which were first detected with RV measurements and then found to undergo transit with subsequent photometric campaigns.

During the first six years of \textsc{Harps} operations, we ran a search for planets orbiting very-low-mass stars on guaranteed time observations. Our sample was composed of $\sim$110 M dwarfs, and we have had success in finding 11 planets \citep[][Delfosse et al. in prep.]{Bonfils:2005b, Bonfils:2007a, Udry:2007, Mayor:2009, Forveille:2009}, although two were actually detected thanks to a complementary sample \citep{Forveille:2011}. Among those detections, we count both the lowest-mass planet orbiting a main-sequence star known to date and the first prototypes of habitable planets \citep[][]{Udry:2007, Mayor:2009}. 

Most recently, we extended that initial sample to more than 300 M dwarfs. For all newly added stars, we focused on the detection of short period planets, with the goal to quickly identify the best candidates for a subsequent transit search. 

In this paper, we report on the first detection obtained with that new sample and strategy, a super-Earth orbiting the M2.5 dwarf \object{GJ\,3634}, and its search for transit with {\it Spitzer} photometry. The paper is structured as follows. In Sect.~\ref{sect:prop} we give the criteria of our extended sample and describe the stellar properties of GJ\,3634. In Sect. \ref{sect:rv} we present and analyze the RVs gathered on GJ 3634. We show their variation are compatible with a planet orbiting the star, plus a long-term drift that is indicative of an additional companion at larger separation. In the same section, we contemplate the possibility that the observed Doppler shifts could be faked by stellar surface inhomogeneities. We searched different stellar activity diagnostics for periodicities, but found no counterpart to the RV variation, which therefore strengthen the planetary interpretation. In Sect.~\ref{sect:phot}, we present our photometric campaign that aimed to search for a possible transit. After considering the detection itself, we take a closer look at the non-detection limit in Sect.~\ref{sect:limit}. Finally, Sect.~\ref{sect:discussion} summarizes our results and discusses the prospects for the RV-educated approach to the search for planets that transit bright nearby stars.

\section{\label{sect:prop}Stellar properties of GJ\,3634}
\object{GJ\,3634} (aka \object{LHS\,2335}) is an M2.5 dwarf \citep{Hawley:1996} seen in the Hydra constellation. It was first referenced by \citet{Eggen:1987} in a catalog of southern high proper motion stars and, according to Simbad, no more than five other times since. \citet{Riedel:2010} recently reported a distance $d=19.8\pm0.6$~pc  ($\pi=50.55\pm1.55~{\rm mas}$) and an apparent brightness V=11.93$\pm$0.01 mag. Together with a declination $\delta=-31.1^{\rm o}$, \object{GJ\,3634} fulfills the criteria of our extended sample, which includes $\sim300$ M dwarfs closer than 20 pc, brighter than V $=$ 12~mag and southward of $\delta = 15^{\rm o}$, as well as $\sim40$ fainter stars kept from our initial sample ($V<14$ mag; $d<11$ pc; $\delta<15^{\rm o}$).

Its infrared photometry \citep[J$=8.361\pm0.023$ mag; K$=7.470\pm0.027$ mag --][]{Skrutskie:2006} and parallax imply an absolute K-band magnitude $M_K=5.99\pm0.16~{\rm mag}$. Using the K-band mass-luminosity relationship of \citet{Delfosse:2000} we attribute a mass $M_\star=0.45~{\rm M_\odot}$ to \object{GJ\,3634}, to which we quote a 10\% uncertainty. We estimate its luminosity $L_\star = 0.025\pm0.004~{\rm L_\odot}$ after converting its absolute K-band magnitude to a bolometric magnitude using its J$-$K color and \citet{Leggett:1992}'s bolometric correction (BC$_{K|J-K} = 2.74\pm0.07$ mag). The metallicity calibrations proposed in the recent years attribute a roughly solar metallicity to GJ\,3634, with [Fe/H]=$-$0.10, $+$0.15 and $+$0.01 dex, following \citet{Bonfils:2005a}, \citet{Johnson:2009a} and \citet{Schlaufman:2010}, respectively. Assuming solar metallicity and an age of 5 Gyr, we also evaluate its radius $R_\star = 0.43\pm0.03 ~{\rm R_\odot}$ from \citet{Baraffe:1998}'s models, with an error estimate combining GJ\,3634's K-band luminosity uncertainty to a model uncertainty of $\sim$5\%. We note that for such a low-luminous star, \citet{Selsis:2007} would place the habitable zone \citep[defined as the region where liquid water can be stable on the surface of a rocky planet;][]{Kasting:1993} at a distance between 0.12 and 0.33 AU. 

\begin{figure}
\includegraphics[width=\linewidth]{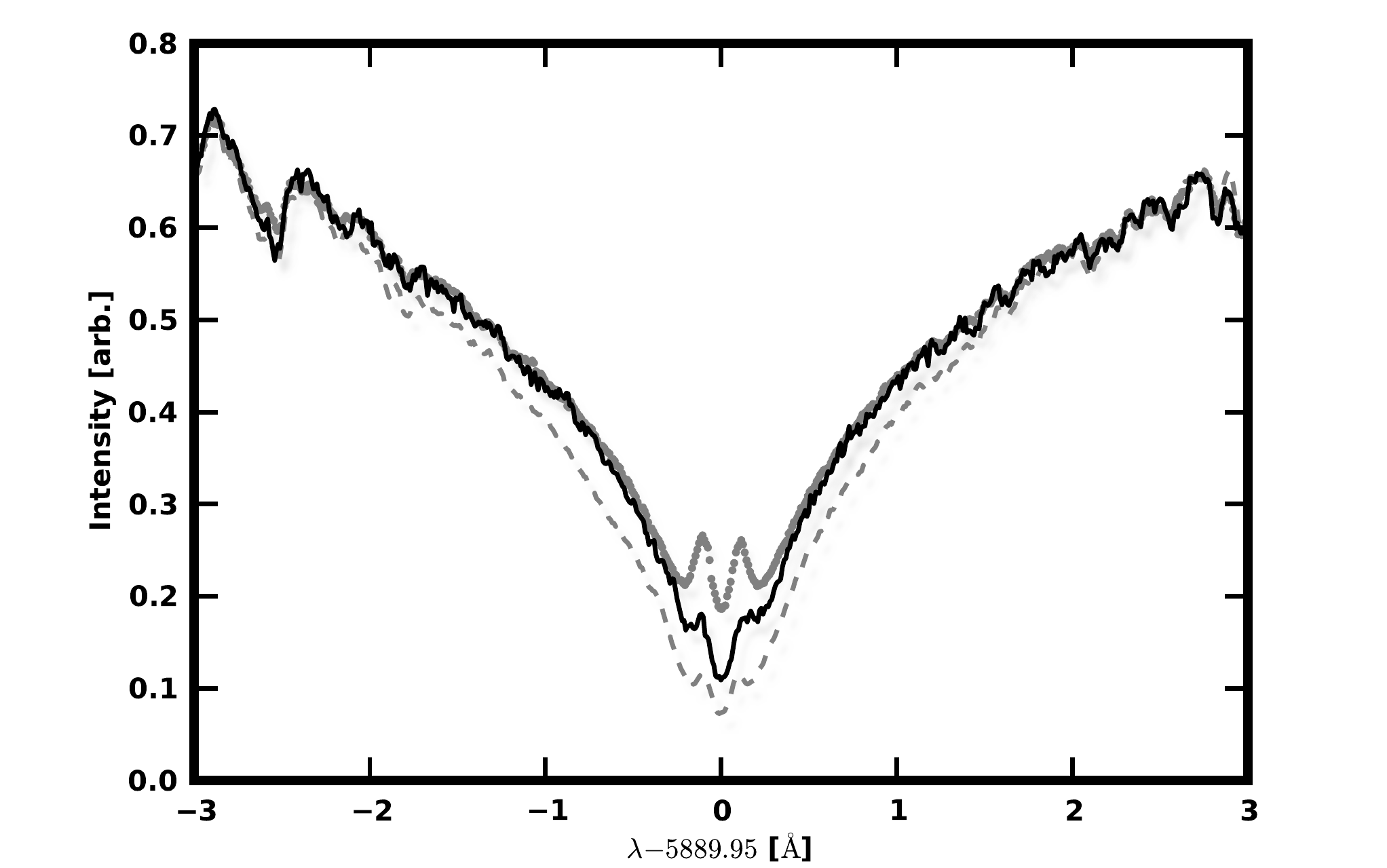}\\
       \caption{Comparison of \ion{Na}{i} D lines for three M2.5 dwarfs, from top to bottom, GJ\,176 (gray dots), GJ\,3634 (black line) and GJ\,581 (gray dashes).}
       \label{fig:naid}
\end{figure}

To assess GJ\,3634's activity level we look at the \ion{Na}{i} doublet. This is known to be an equivalent diagnostic to \ion{Ca}{ii} H\&K emission lines and a more adequate choice when blue-most spectral orders have low signal-to-noise ratio \citep{Diaz:2007, Silva:2010}. We compare GJ\,3634 to a quiet \citep[GJ\,581 --][]{Bonfils:2005a} and a moderately active \citep[GJ\,176 --][]{Forveille:2009} M dwarf, and diagnose an intermediate level of activity (see Figure~\ref{fig:naid}). Also, its galactic velocities \citep[$U=-35, V=-25, W=-20$ km/s --][]{Hawley:1996} place GJ\,3634 in a position between the young and old disks populations \citep{Leggett:1992}. Stars of this population have a probable age $>3$ Gyr \citep{Haywood:1997}, which is consistent with the level of activity we estimate from the  \ion{Na}{i} doublet. Together with the low $v \sin i \lesssim 1$~km/s  we estimate from \textsc{Harps} spectra, we expect that GJ\,3634's magnetic activity is too low to affect our radial-velocity measurements, at least on short time scales.

\begin{table}
\caption{
\label{table:stellar}
Observed and inferred stellar parameters for GJ~3634}
\begin{tabular}{l@{}lc}
\hline
 \multicolumn{2}{l}{\bf Parameters}
& \multicolumn{1}{c}{\bf GJ~3634} \\
\hline
$\alpha$  & & $10^{\rm h}58^{\rm m}35.10^{\rm s}$ $^\dagger$ \\
$\delta$  &  & $-31^{\rm o}08^{\rm '}39.1^{\rm ''}$ $^\dagger$\\
Spectral Type   &                & M2.5\\
$\pi$           &[mas]          & $50.55 \pm 1.55$ \\
d                 & [pc]             & $19.8\pm0.6$\\
V                       &  [mag]       & $11.93\pm0.01$ \\
$M_V$           &    [mag]            & $10.45 \pm 0.15$ \\
K                       &   [mag]             & $7.470 \pm 0.027$\\
$M_K$           &     [mag]           & $5.99 \pm 0.16$\\
$L_\star$       & [${\rm L_\odot}$]          &  $0.020\pm0.002$\\
$M_\star$       & [${\rm M_\odot}$]             & $ 0.45\pm0.05 $\\
$R_\star$          & [$\rm R_\odot$]    &  $0.43\pm0.03$\\
\hline
\end{tabular}
\\$^{\rm \dagger}$ \citet{Bakos:2002}
\end{table}

\section{\label{sect:rv}Radial-velocity detection}
\subsection{\label{subsect:rv1}Data and orbital analysis}
\begin{figure}
\includegraphics[width=\linewidth]{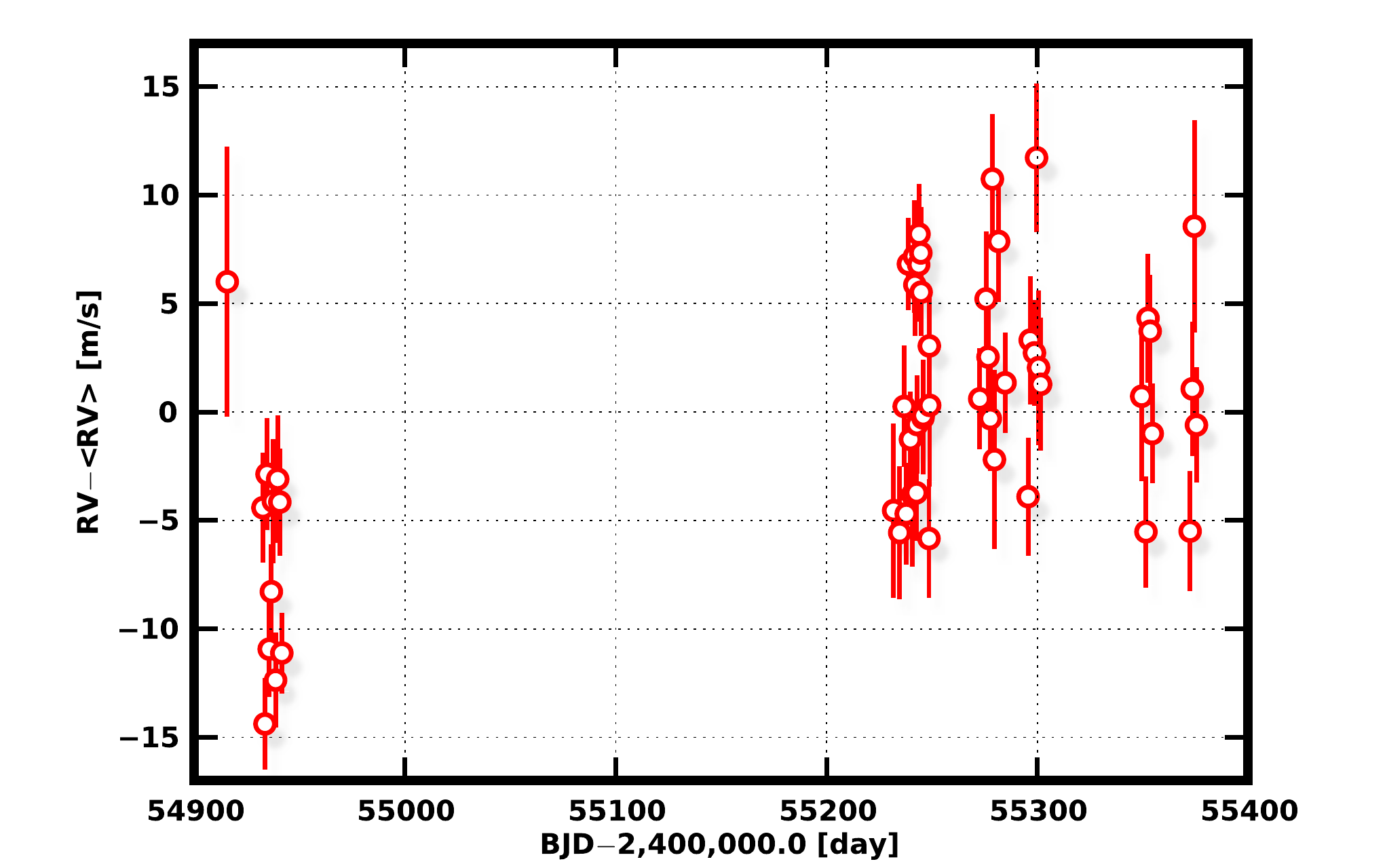}\\
\includegraphics[width=\linewidth]{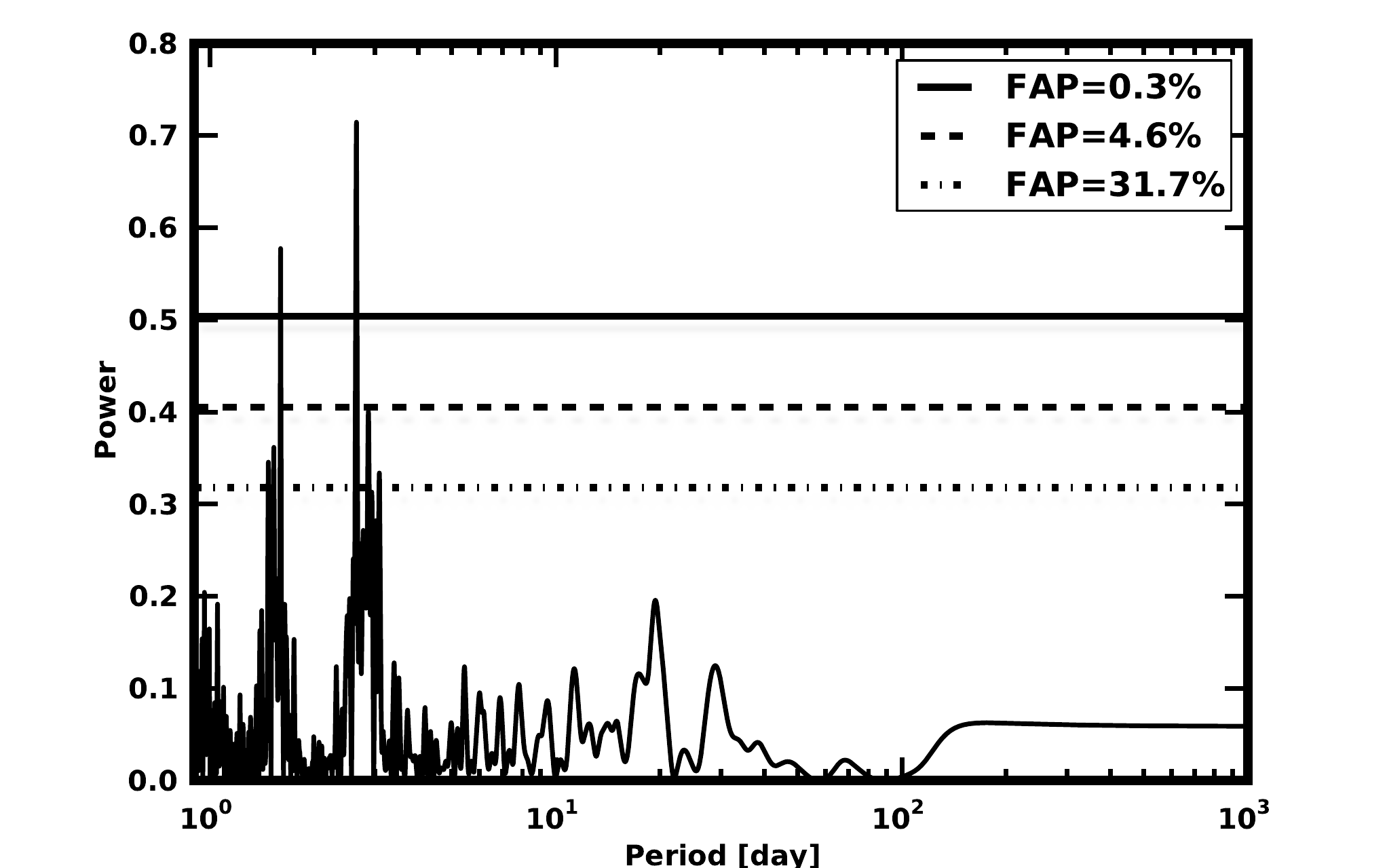}
       \caption{{\it Upper panel:}  RV time series of GJ\,3634, collected in 2009 (Barycentric Julian Date$<$2,455,198) and in 2010 (BJD$>$2,455,198). {\it Bottom panel:} Periodogram of GJ\,3634 RVs collected in 2010. The horizontal lines show different levels of false-alarm probabilities.}
       \label{fig:fig2}
\end{figure}

We observed \object{GJ\,3634} with the \textsc{Harps} spectrograph, the state-of-the-art velocimeter fiber-fed by the ESO/3.6-m telescope \citep{Mayor:2003, Pepe:2004}. Our setting remains the same as in our GTO program and we refer the reader to \citet[][sub.]{Bonfils:2011} for a detailed description. 

We started GJ\,3634 observations by a single 300-s exposure on 2009 March 25. We verified that the target was neither a double-line spectroscopic binary, nor a fast rotator, and that the precision was sufficient for a planet search. Among other stars selected with a single measurement, we re-observed \object{GJ\,3634} about two weeks later with an exposure of 900 s. Sometimes, the two weeks lap time enables the identification of single-line spectroscopic binaries right after the second measurement, in which case we would discard the star from the target list. For GJ\,3634 however, we continued the observations at a pace of one measurement per night during 10 consecutive nights, and with 900-s exposures.

After those observations, the dispersion of GJ\,3634 RVs ($\sigma_e=5.8~{\rm m/s}$) appeared to be in excess of their estimated uncertainties ($<\sigma_i>=2.6~{\rm m/s}$). Power excess could already be seen in the periodogram of those velocities at the periods 1.6 and 2.7 day (one-day aliased of each other), albeit with a modest false-alarm probability (FAP) of $\sim4.3$\%. GJ\,3634 nevertheless became a high-priority target for the next season and this year, we gathered 43 more measurements, which makes a total of 54 (Fig.~\ref{fig:fig2} and Table 2$^\star$). \addtocounter{table}{1}

A periodogram of this season's velocities shows a clear power excess ($p_0=0.71$) around a 2.65-day period, plus a less powerful pic at the sidereal-day alias of this period, 1.60 day ($p=0.58$). Shuffling the RVs but retaining the dates, we created 10,000 virtual data sets and computed their periodograms. The distribution of their maxima has a mean value of 0.29, with a standard deviation of 0.06. None of the maxima measured in the simulated periodograms is as high or higher than the power maxima measured on the  periodogram of the original data, which suggests a FAP lower than $O(1/10,000)$. We also computed an analytical estimate of the FAP with \citet{Cumming:2004}'s prescription: $FAP  \simeq M.(1-p_0)^{(N-3)/2}$, where $M$ is the number of independent frequencies in the periodogram, $p_0$ its highest power value and $N$ the number of measurements. We approximate $M$ by the inverse of the time span of our observations and obtain the extremely low $FAP$ value of $\sim 2.5 \times 10^{-9}$.

Still considering velocities from 2010 only, we performed a Keplerian fit, using the period of the detected signal as a starting guess. Our minimization converges toward a solution with a reduced $\chi^2_{\rm red}=0.95\pm0.23$, greatly improved compared with the reduced $\chi^2_{\rm red}=3.14\pm0.39$ of a fit by a constant. 

Now considering all data from both 2009 and 2010, we subtracted that best fit to all velocities. We find that the RVs of 2009 have an average value $\sim8$ m/s lower compared with those of 2010. This RV offset most likely betrays the presence of an additional companion around GJ\,3634, though more data are required to complete the orbit and confirm this interpretation. The periodogram of all RVs taken together is actually dominated by the power of that long-term variation and, for a better legibility, we chose to restrict our periodogram analysis to the subset of the 2010 RVs. 

Finally, we found that a planet plus a quadratic drift is a good model to describe all RVs. We use this model together with a Markov Chain Monte Carlo (MCMC) algorithm to perform a Bayesian analysis of all RVs \citep[e.g.][]{Gregory:2005, Gregory:2007, Ford:2005}. For each parameter, we marginalize over all other parameters, take the median of the posterior density function as the optimal value, and the centered 68\% interval as an error estimate. Our model was composed of nine parameters : the orbital period ($P=2.6459\pm 0.0006$~day), the semi-amplitude ($K_1=5.60\pm0.57~{\rm m/s}$), the time of passage at periastron ($T=2,454,917.04^{+0.82}_{-0.52}$), the orbital eccentricity ($e=0.09_{-0.06}^{+0.09}$), the argument of periastron ($\omega=100\pm71^{\rm o}$), the slope ($slope = 21.1\pm2.8$ m/s/yr) and quadrature ($quad = -10.3\pm2.4 {\rm m/s/yr}^2$) of the long-term drift, and a jitter component ($\epsilon_j=0.47_{-0.35}^{+0.51}$ m/s), quadratically co-added to the photon noise. We note that both $e$ and $\epsilon_j$ are compatible with zero and give upper values $e<0.31$ and $\epsilon_j<1.8$ m/s, valid with a 99\% confidence level. 

The Bayesian approach also offers the possibility to evaluate the confidence level of our detection more rigorously. We thus ran MCMC chains for several models and computed their relative Bayes factor. We found that a model composed of {\it 1 planet $+$ a quadratic drift} is favored over a model composed of {\it a single planet} by a factor $\sim 5\times 10^5$. And over a constant model (i.e. no planet and no drift), we found that  {\it 1 planet $+$ a quadratic drift} model is favored by a factor of $\sim 1 \times 10^9$, which is in line with the very strong detection.

\begin{figure}
\centering
\includegraphics[width=\linewidth]{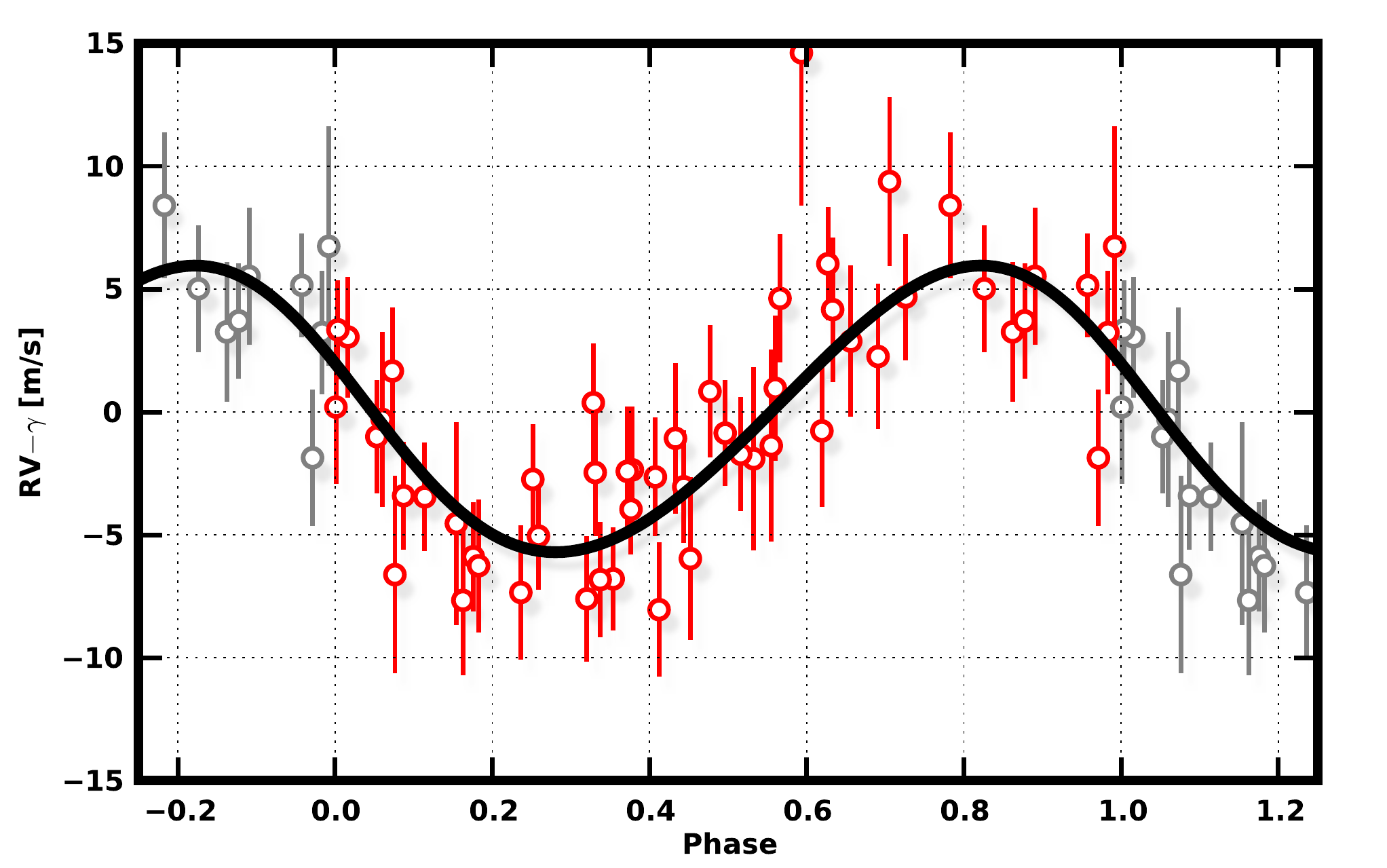}
\includegraphics[width=\linewidth]{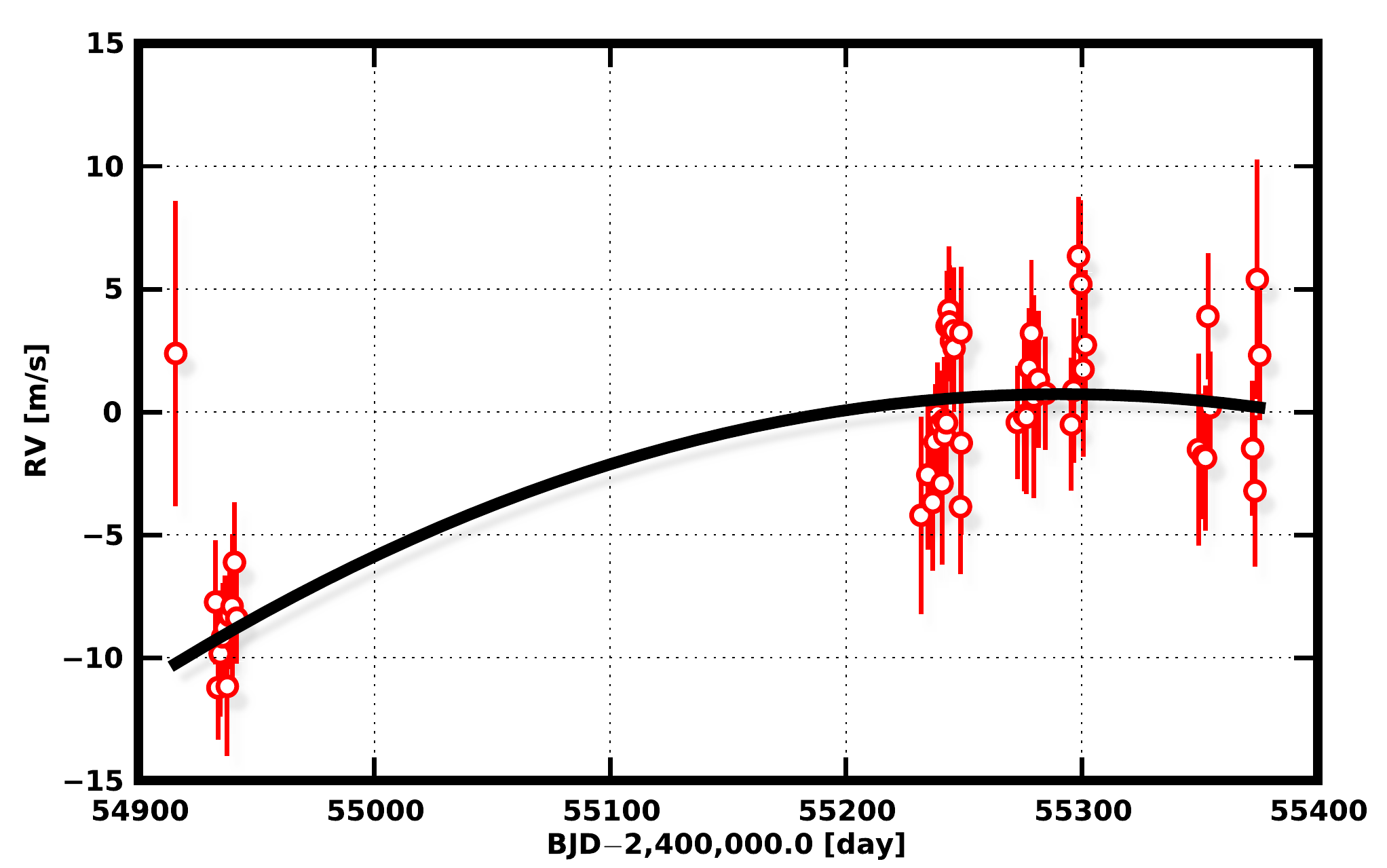}
       \caption{Decomposition of our nominal orbital model for GJ3634. The top panel shows RVs with the contribution of the long-term drift subtracted and phased with the planet's orbital period. The bottom panel shows RVs as a function of time, with the Keplerian contribution removed. Gray points duplicate some of the red points. The model is also decomposed and plotted with a solid curve in each panel.}
       \label{fig:orb}
\end{figure}

Finally, we show the optimal solution in Fig..~\ref{fig:orb}, and using $M_\star = 0.45\pm0.05~{\rm M_\odot}$ we convert the orbital parameters into a planetary minimum mass $m \sin i = 7.05\pm0.89~{\rm M_\oplus}$.

\subsection{Activity diagnostics}
We compute the RVs of GJ\,3634 by measuring the Doppler shifts of spectra recorded by the \textsc{Harps} spectrograph. At first, the method assumes that the spectrum emerging from the stellar surface remains unchanged over time, except for the Doppler shifts attributed to one or more perturbing bodies that impose velocity changes on the star. Then, one has to consider the stellar phenomena able to alter the emergent spectrum. Spots and plages, for instance, can break the balance between the blue- and red-shifted halves of a rotating star. As the star rotates, the stellar inhomogeneities modulate the overall integrated spectrum and bias RV measurements. In some cases, the modulation can even mimic the Keplerian wobble expected from an orbiting planet, like for GJ\,674 \citep{Bonfils:2007a}. In our search for planets orbiting M dwarfs, the RV modulations we identified were actually more often caused spots or plages than planets (Bonfils et al. 2011, submitted). 

Although GJ\,3634's low v sin i ($\lesssim 1$ km/s) makes correspond an improbable inclination to a few-day rotational period, we applied several diagnostics to distinguish between stellar activity and true RV shifts. A first class of diagnostics is based on the spectral line asymmetry. We measure both the bisector-inverse slope (BIS) and the full-width at half-maximum (FWHM) of the cross-correlation function (an averaged spectral line of \textsc{Harps} spectra). None seems to show any periodicity, nor a correlation with the radial velocities. This type of analysis however looses its power for low $v sin i$. Alternatively, a second class of diagnostics is based on the photometric and spectral signatures of active regions of the stellar surface. Spots, plages, and filaments produce back emission in \ion{Ca}{ii} H\&K, \ion{Na}{I}{D} and H$\alpha$ lines. We therefore also investigated the possible variation of indices based on these lines, but fund no counterpart to the observed RV variation. 

\section{\label{sect:phot}Transit search with {\it Warm Spitzer}}
The {\it a priori} geometric probability that GJ\,3634b transits its host star is $\sim R_\star/a\simeq 7\%$. We therefore decided to follow-up on the detection of GJ\,3634b with a photometric campaign, and part of our RV observations aimed at refining the orbital ephemeris to narrow the probable transit-search window. 

\subsection{Light curve}

For a rocky composition, the transit depth could be shallower than 1 mmag, making a ground-based transit detection extremely difficult. We therefore opted for the Infra-Red Array Camera (IRAC) onboard {\it Warm Spitzer} \citep{Soifer:2007, Fazio:2004} and scheduled our observations to cover $\sim$2$\sigma$ of the probable transit window, from 2010 July 12 17h50 UT to 2010 July 13 00h15 UT. We chose to observe in the 4.5-$\mu$m channel because it exhibits the lowest intrapixel-sensitivity variation. Indeed, combined with the low-frequency jitter of the telescope, this inhomogeneous response produces a strong correlation between the recorded flux and the position of the star on the detector (Knutson et al. 2008 and references therein). We used the established technique of continuous staring in non-dithered subarray mode with the longest exposure time for which the star would not be saturated on the detector (0.32s). 

\begin{figure}
\centering
\includegraphics[width=\linewidth]{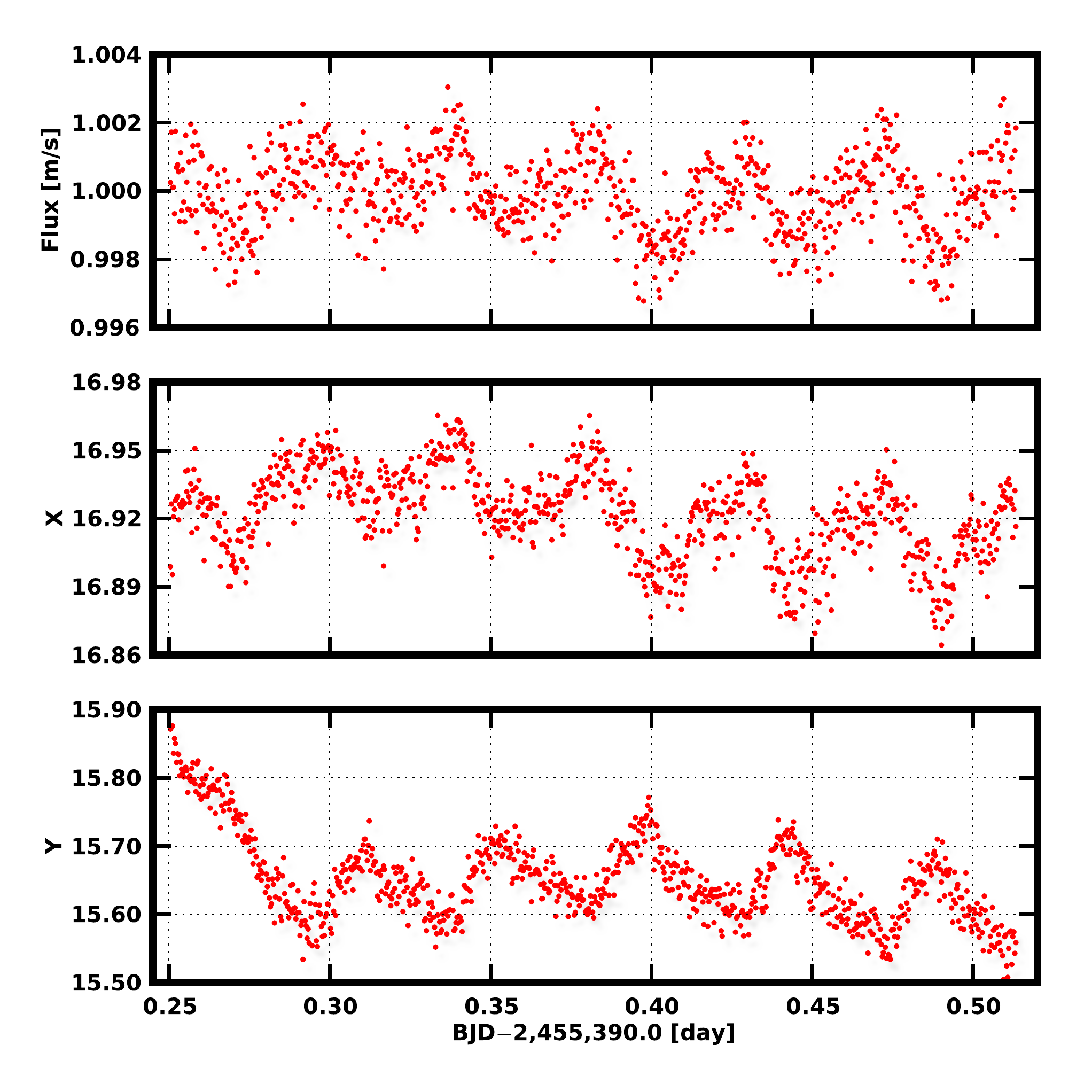}
\caption{$Top$: Raw 4.5$\mu$m {\it Warm Spitzer} light curve for GJ\,3634. $Middle$ and $bottom$: evolution of $x$- and $y$-positions of the PSF center during the run.}
       \label{fig:MG1}
\end{figure}

Our data consist of 845 blocks of 64 individual subarray images. We used the Basic Calibrated Data (BCD) produced by the {\it Spitzer} standard pipeline (version S18.18.0) and, after conversion from specific intensity (MJy/sr) to photon counts, we obtained aperture photometry in each subarray image using the {\tt IRAF/DAOPHOT}\footnote{{\tt IRAF} is distributed by the National Optical Astronomy Observatory, which is operated by the Association of Universities for Research in Astronomy, Inc., under cooperative agreement with the National Science Foundation.}  software \citep{Stetson:1987}. We determined the stellar position by fitting a Gaussian profile to the stellar image, and obtained our best results with an aperture of 3.5 pixels. In each image we subtracted a mean sky-background measured in an annulus extending from 12 to 15 pixels from the aperture center. For each block of 64 subarray images, we discarded the discrepant values for the measurements of flux, background, and the $x$- and $y$-positions using a 10-$\sigma$ median clipping, and averaged the remaining values. Only 0.3 $\%$ of the measured fluxes were discarded. To estimate the error on the averaged value we chose to divide the r.m.s. of the block by the squared root of the number of measurements we kept. Figure~\ref{fig:MG1} shows the evolution of the measured flux and position time series. It clearly shows the correlation of the photometry with the stellar position, which leads to a correlated noise at the level of a few mmags in the light curve. It also shows that the flux decreased sharply during the first 5 minutes of the run, and this drop is correlated with a sharp variation of the $y$-position. We decided to discard these first 5 minutes (13 measurements) from our analysis, and accordingly obtained a final light curve with 832 measurements. 

\begin{figure}
\centering
\includegraphics[width=\linewidth]{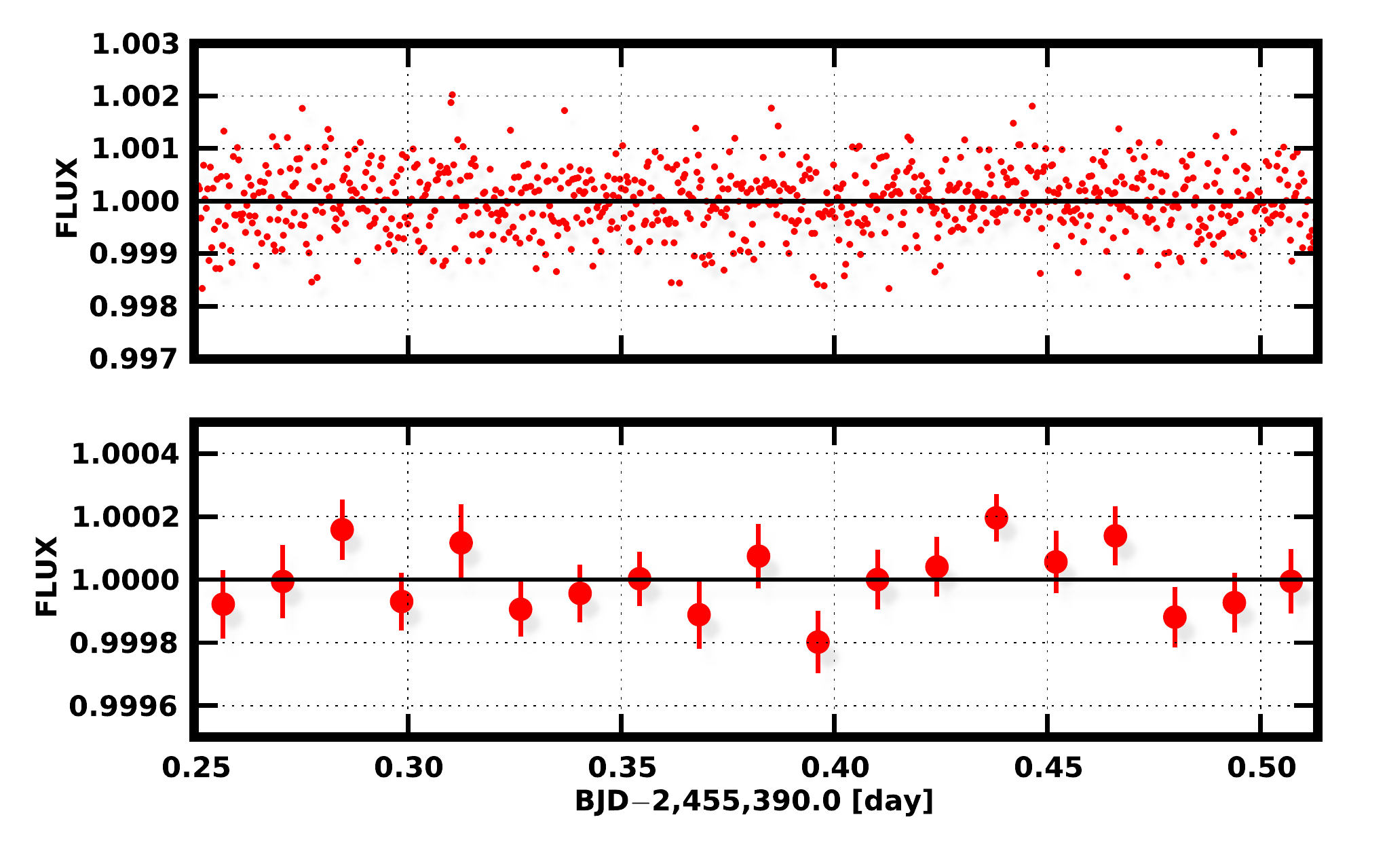}
\caption{$Top$: 4.5 $\mu$m light curve detrended by the baseline function $A(dx, dy, dt)$ (see Eq. 1). $Bottom$: same light curve binned per 20 min time bins.}
       \label{fig:MG2}
\end{figure}

The flux-position correlation is well described by a quadratic polynomium in $x$ and $y$ :
\begin{equation}
A(dx, dy, t) = a_1+ a_2 dt + a_3 dx +a_4 dx^2+a_5 dy+a_6 dy^2 +a_7 dxdy \textrm{,}
\end{equation}
where $dx$ and $dy$ are the distance of the PSF center to the center of the pixel, and $dt$ is the time elapsed since 15 min before the start of the run.
Detrending the light curve by this seven-parameter function leads to a time-series nearly free of correlated noise (see Fig.~\ref{fig:MG2} and Table 3$^\star$). The r.m.s. is 611 ppm, much similar to the mean individual error ($\sim$ 616 ppm) and once binned per 20 minutes, the r.m.s. decreases to 100 ppm. \addtocounter{table}{1}

\subsection{Orbital analysis}

To determine the $posterior$ transit probability of GJ\,3634b, we pooled the radial-velocity and the photometry and applied another Bayesian analysis using MCMC. This analysis closely follows those presented for HD\,40307b and CoRoT-2b \citep{Gillon:2010b, Gillon:2010c}, and we refer the reader to these for details regarding the choice of parameters and their priors. To briefly recall, the model includes a planet on a Keplerian orbit, a quadratic RV drift, a photometric baseline, and the stellar mass and radius. At each step of the Markov chain a set of parameters is proposed and, to accept (or reject) the proposal, we evaluate the joint likelihood of the data (RV$+$photometry) and the parameters. The planet's parameters act on the radial-velocity likelihood and, if transiting, on the photometry likelihood. The parameters of the quadratic drift  and the stellar mass affect only the RV likelihood, and the baseline and the stellar radius only the photometry likelihood.

To improve the efficiency of convergence, we decoupled the linear parameters (the photometric baseline, the RV trend, and the gamma velocity) from the MCMC and fitted them with a least-square minimization. More precisely, the quadratic drift counts two parameters (the $slope$ and the $curvature$) and the photometric baseline counts seven (Eq. 1). We also describe the planet with the following eight parameters: the orbital period $P$, $\sqrt{e} \cos{\omega}$ and $\sqrt{e} \sin{\omega}$, where $e$ and $\omega$ are the eccentricity and argument of periastron, a parameter $K_2 = K  \sqrt{1-e^2}   \textrm{ } P^{1/3} $ to replace the RV semi-amplitude $K$, the time $T_{tr}$ (when the true anomaly  $\nu = \pi/2 - \omega$), the impact parameter  $b' = a \cos{i}/R_\ast$ and the planetary radius $R_p$ chosen between 1.2 $R_\oplus$ for a pure-iron planet \citep{Seager:2007} and an arbitrary upper limit of 12~$R_\oplus$. The stellar mass and radius are the only parameters with non-uniform priors; they are drawn randomly at each step of the Markov chain following their normal distributions, $N(0.45,0.05^2)$ ${\rm M_\odot}$ and $N(0.43,0.03^2)$ ${\rm R_\odot}$. Note finally that we chose not to include a description of the stellar limb darkening, nor of the Rossiter-McLaughlin anomaly, because both are expected to be small.

\begin{figure*}
\centering
\includegraphics[width=\linewidth]{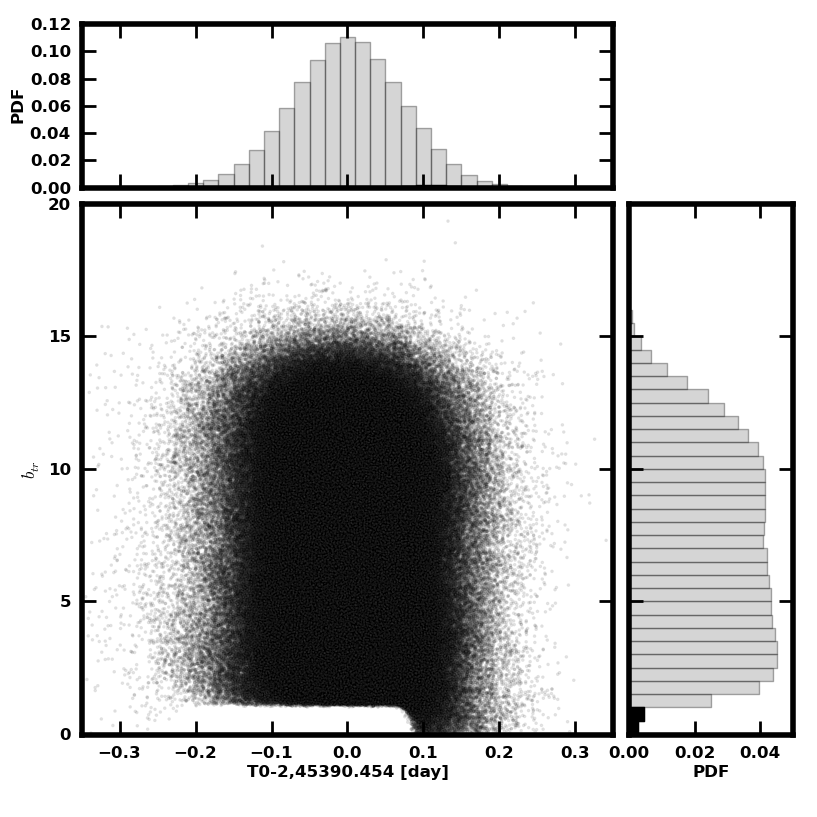}
 \caption{Joint marginal posterior distribution of $T_0$ (transit timing) and $b_{tr}$ (transit impact parameter) = $(a\cos{i}/R_\ast)[(1-e^2)/(1+e\sin(\omega)]$. Full transit configurations represent 0.5\% of the PDF and are shown in black in the histograms.} 
       \label{fig:MG3}
\end{figure*}

We ran the analysis in two steps. First, we performed a Markov chain of  500\,000 steps to assess the level of correlated noise in the photometry and the jitter noise in the RVs. We found that no jitter noise is required for the RVs while we multiplied the photometric error by $\beta_{\rm \textrm{} red} = 1.07$ \citep[e.g. ][]{Gillon:2010c}. Then, we performed a new chain of 500\,000 steps with updated errors. Our MCMC converges on a statistical description of possible solutions. We marginalized the posterior distribution of each parameters and calculated their median values and 68.3\% intervals (Table~\ref{tab:params}). Among the solutions, none corresponds to a detected transit, and among all configurations with a transit, 92\% are rejected. Only few transit configurations remain unexplored (Fig.~\ref{fig:MG3}), and the posterior transit probability decreases to 0.9\% (and even to 0.5\%  if only total transits are retained). 

This global analysis also improves the measurement of orbital parameters. Indeed, the small subset of transit configurations that are inconsistent with the photometry removes as many orbital inclinations --and therefore planetary masses-- from the possible solutions. It thus provides a statistical description for the true mass of GJ\,3634b ($m_p=8.4^{+4.0}_{-1.5} {\rm M_\oplus}$).

\begin{table}
\begin{center}
\begin{tabular}{lcccl}
\hline
Parameter  & Value + errors & Units \\ \noalign {\smallskip}
\hline \noalign {\smallskip}
Jump parameters &   &  \\ \noalign {\smallskip}
\hline \noalign {\smallskip}
$dF = (R_p/R_\ast)^2$ & $0.020 \pm 0.012$ &   \\ \noalign {\smallskip}
$ b' = a \cos(i) / R_\ast$ & $7.5 \pm 4.3$  &   $R_*$  \\ \noalign {\smallskip}
Transit epoch  $ T_0$ - 2450000 & $5390.454\pm0.073$   &  BJD  \\ \noalign {\smallskip}
Orbital period  $ P$ & $2.64561\pm0.00066$ &    days  \\ \noalign {\smallskip}
$K_2 = K  \sqrt{1-e^2}   \textrm{ } P^{1/3}$ & $7.68 \pm 0.75$ &    \\ \noalign {\smallskip}
$\sqrt{e}\cos{\omega}$ & $-0.04^{+0.20}_{-0.19}$ & \\ \noalign {\smallskip}
$\sqrt{e}\sin{\omega}$  & $0.17^{+0.19}_{-0.24}$ & \\ \noalign {\smallskip} 
\hline \noalign {\smallskip}
RV baseline parameters &   &  \\ \noalign {\smallskip}
\hline \noalign {\smallskip}
Systemic RV & $5.18635^{0.00015}_{-0.00016}$  &  km/s  \\ \noalign {\smallskip}
Slope & $21.77 \pm 0.77$ &   m/s/yr  \\ \noalign {\smallskip}
Curvature & $-11.62 \pm 0.77$ &   m/s/yr$^2$  \\ \noalign {\smallskip}
\hline \noalign {\smallskip}
Deduced parameters &    & &  \\ \noalign {\smallskip}
\hline \noalign {\smallskip}
RV $K$ & $5.59 \pm 0.55 $ &  m/s   \\ \noalign {\smallskip}
Orbital semi-major axis $a$ & $ 0.0287^{+0.0010}_{-0.0011}$ &AU \\ \noalign {\smallskip}
Orbital inclination $i$ & $59^{+18}_{-24}$  & deg \\ \noalign {\smallskip}
Orbital eccentricity $e$ & $0.080^{+0.094}_{-0.057}$ &  \\ \noalign {\smallskip}
Upper limit to $e$ (99\%)& $<0.36$ & \\ \noalign {\smallskip}
Argument of periastron $\omega $ & $84^{+52}_{-130}$   & deg \\ \noalign {\smallskip}
$M_p \sin{i}$ & $7.0^{+0.9}_{-0.8}$ & $M_\oplus$ \\ \noalign {\smallskip}
$M_p $ & $8.4^{+4.0}_{-1.5}$ & $M_\oplus$ \\ \noalign {\smallskip}
\hline \noalign {\smallskip}
Eclipse probabilities&    & &  \\ \noalign {\smallskip}
\hline \noalign {\smallskip}
${\rm P_{transit}}$ & 0.9\% &  \\ \noalign {\smallskip}
${\rm P_{fulltransit}}$ & 0.5\% &  \\ \noalign {\smallskip}
${\rm P_{occultation}}$ & 0.7\% &  \\ \noalign {\smallskip}
${\rm P_{fulloccultation}}$ & 0.4\% &  \\ \noalign {\smallskip}
\hline \noalign {\smallskip}
\end{tabular}
\caption{\label{tab:params}Median and 1-$\sigma$ limits of the marginal posterior distributions of the orbital parameters.}
\end{center}
\end{table}

\section{\label{sect:limit}Residuals and detection limits}
At this point, we identified the Keplerian motion induced by a super-Earth, searched for its possible transit and were able to reject most transit configurations. We also identified a second signal that most probably corresponds to the incomplete orbit of a companion of whose true nature is yet unknown. The data of 2010 are weakly sensitive to that signal, suggesting an orbital period greater than their time span ($>200$ d), and a mass $\gtrsim 32~{\rm M_\oplus}$ ($\sim 2~{\rm M_{Nep}}$).

\begin{figure}
\centering
\includegraphics[width=\linewidth]{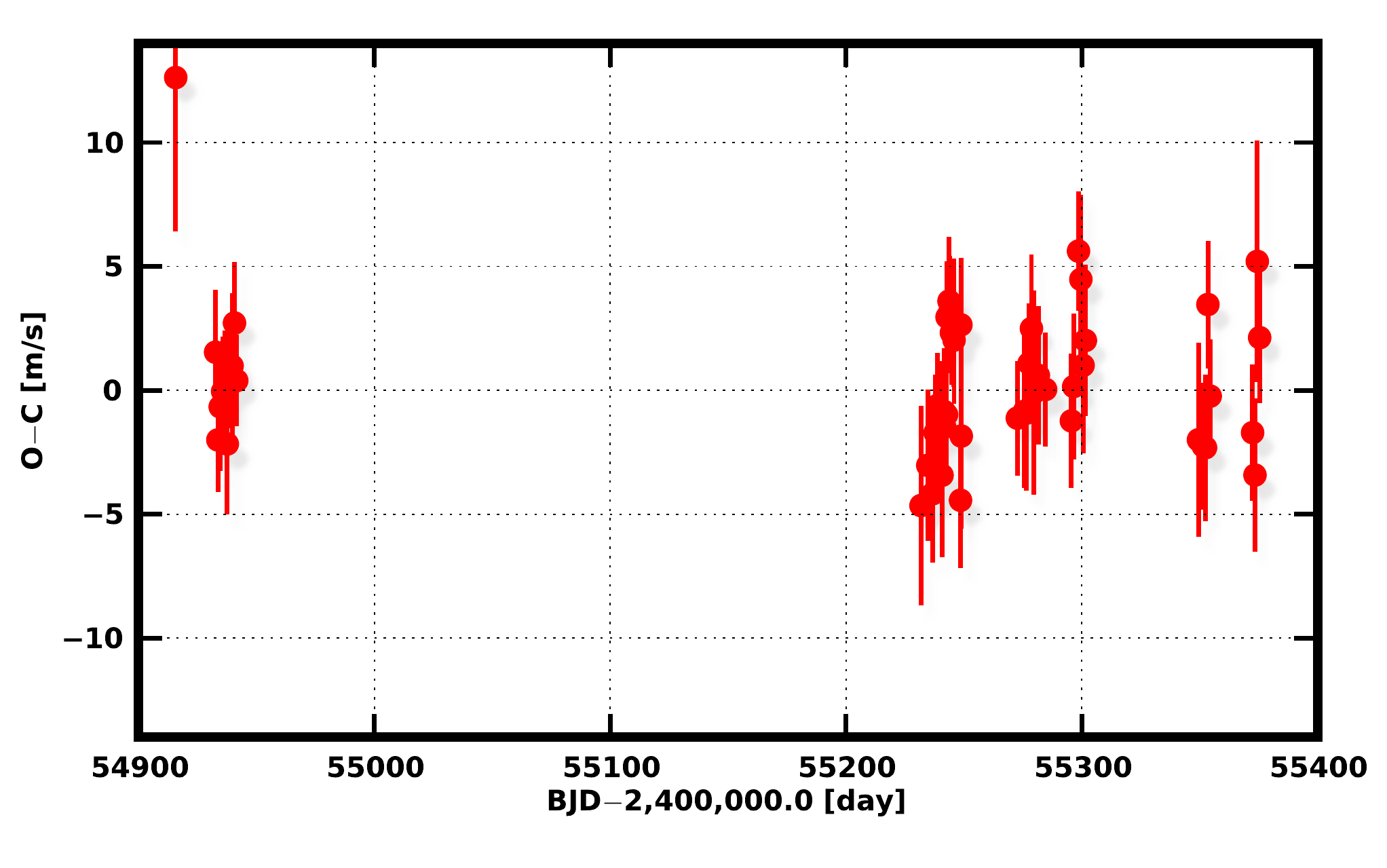}
\includegraphics[width=\linewidth]{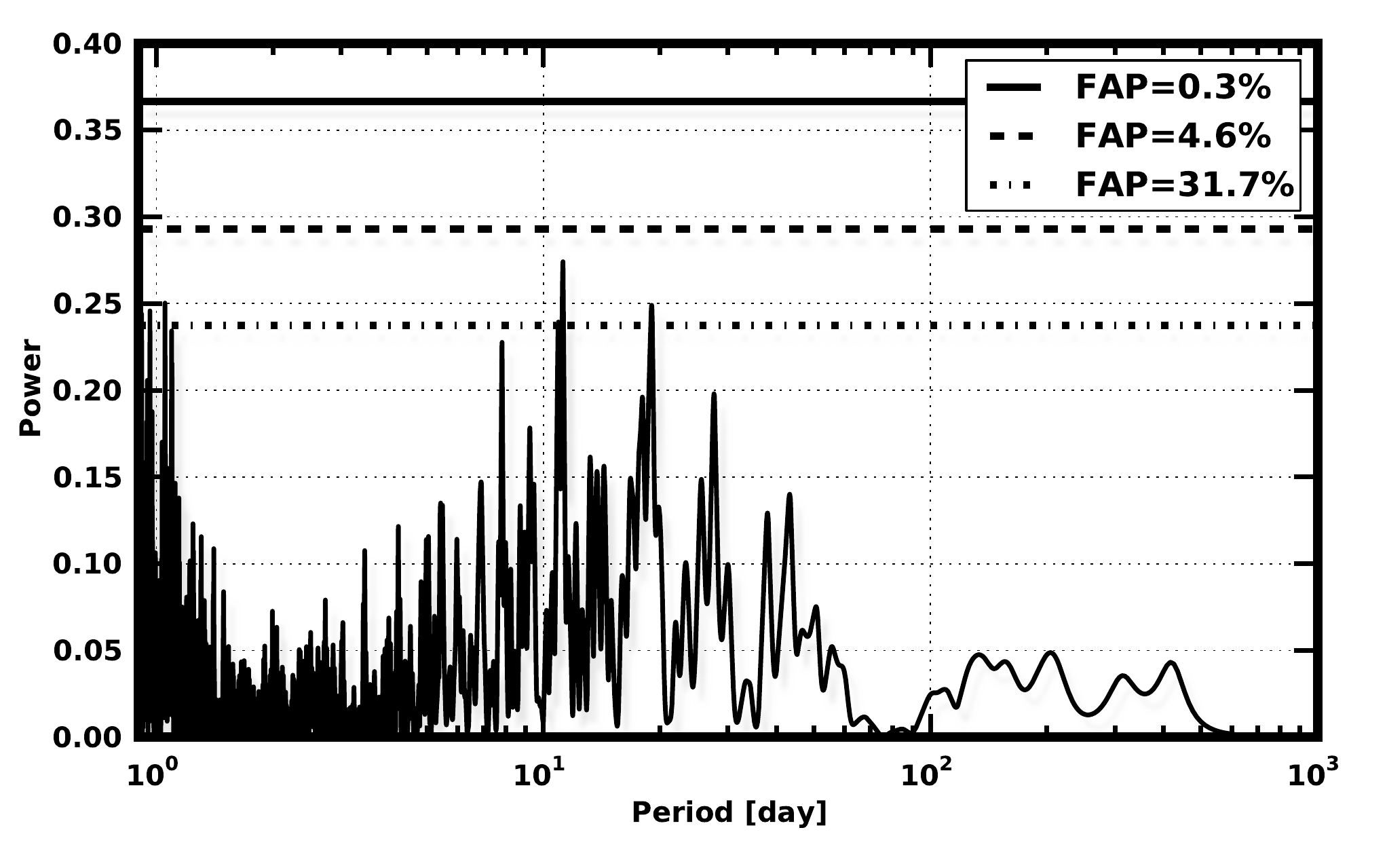}
       \caption{{\it Upper panel:} RV residuals around the best solution for a {\it 1 planet $+$ quadratic drift} model  {\it Bottom panel:} Periodogram of the residuals.}
       \label{fig:residuals}
\end{figure}

The habitable zone of GJ3634 is located between $\sim$0.12 and $\sim$0.33 AU from the star, corresponding to orbital periods from $\sim$22 to 104 d. There is therefore much interest in characterizing our sensitivity in the period range below 200 d. Figure~\ref{fig:residuals} shows the RV residuals once the best fit for the {\it 1 planet $+$ quadratic drift} model has been removed. They have a dispersion $\sigma=2.32$ m/s. The reduced $\chi^2=1.00\pm0.10$ of the solution suggests the remaining dispersion is explained by the measurement uncertainties and argues against more complex models. The periodogram of the residuals (Fig.~\ref{fig:residuals}, bottom panel) does not display significant power excess. The most important pic however is located at a period $\sim$19 d (FAP=10\%), close to the habitable zone's inner edge, and will retain our attention when more RVs will be collected. 

\begin{figure}
\centering
\includegraphics[width=1.\linewidth]{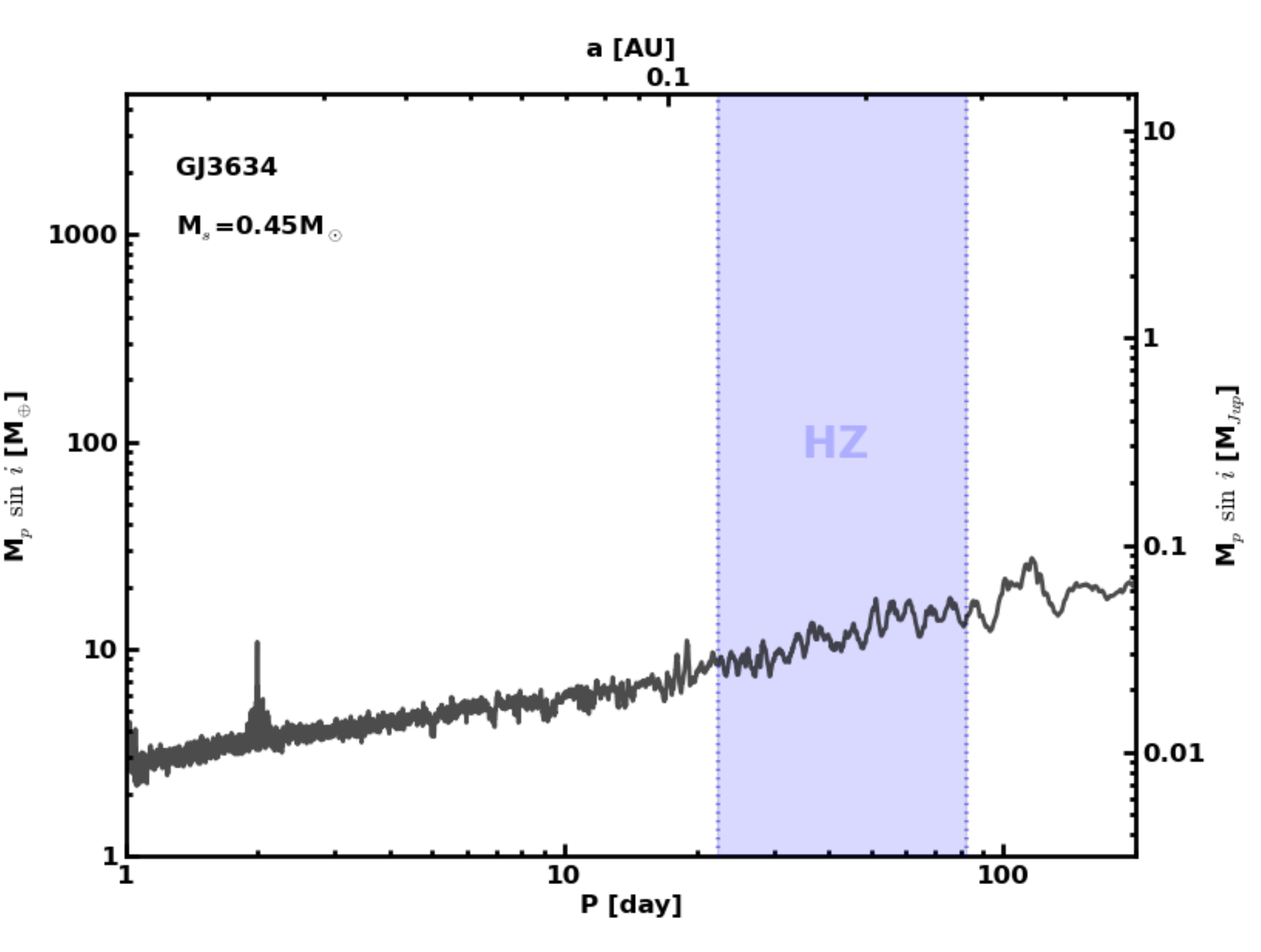}
       \caption{Detection limit imposed by RVs residuals around the best solution for a {\it 1 planet $+$ quadratic drift} model, for periods $<$200 d. The light blue area delineates the habitable zone, using Venus and early-Mars criterions \citep{Selsis:2007}.}
       \label{fig:Limit}
\end{figure}

We then turn to assess which planets, as a function of minimum mass and orbital period, are rejected given the residuals. Because the model to describe the long-term variation is uncertain, we restrict our analysis to periods $\le$200 d. We use a method applied by \citet{Cumming:1999, Cumming:2008} and \citet{Zechmeister:2009b}, which we recently employed to derive detection limits of 85 M dwarfs from our sample (Bonfils et al., submitted). In brief, we carried out bootstrap resampling of the residuals, generated 1,000 virtual data sets, and computed their periodograms. We used the 1,000 periodograms to build an empirical power distribution, scan the periods and determine the threshold that encompasses 99\% of the power realizations. We then injected faked circular orbits in the observed residuals (with 12 different phases), and increased their semi-amplitude until they produced a power as high or higher than our power threshold in the periodogram (for all trial phases). This semi-amplitude is the level above which a planet can be conservatively rejected (with a confidence level of 99\%). Finally, we converted this limit in a minimum-mass limit using Kepler's law and our estimate of the stellar mass (Fig.~\ref{fig:Limit}). To aid the reader, we also report in Fig.~\ref{fig:Limit} the putative habitable zone, following the \citet{Selsis:2007}'s prescription.

The detection limit shows that we rule out additional planets more massive than $>10~{\rm M_\oplus}$ up to periods well above 10~d, except for a very narrow period range around $2$~d, where our sensitivity decreases because of the observation sampling. In GJ3634's habitable zone, we exclude planets more massive than $m\sin i\sim8-20~{\rm M_\oplus}$, from the inner to the outer edges.

\section{\label{sect:discussion}Conclusion}

We report the detection of a $m\sin i = 7.0^{+0.9}_{-0.8}~{\rm M_\oplus}$ planet orbiting the nearby M dwarf GJ\,3634. We followed-up on the RV detection with IRAC/{\it Spitzer} photometry to check whether the planet could be seen to transit its parent star. Our light curve confidently rejects most of the orbital configurations corresponding to a transiting planet as the posterior probability for full transit is decreased to $\sim 0.5\%$.

That detection adds to the handful of low-mass Neptunes and super-Earths detected so far. About a dozen planets are known with $m \sin i \lesssim 8~{\rm M_\oplus}$ and almost half orbit M dwarfs. Among those, the transiting GJ\,1214b has a similar mass \cite[$m \sin i = 6.55\pm0.98 {\rm M_\oplus}$ --][]{Charbonneau:2009} as GJ\,3634b. GJ\,1214b is found to have a thick atmosphere and, in structure and composition, resembles more a Neptune-like planets than a large rocky planet. It has also been suggested that, if this characteristic is shared by habitable super-Earths detected in RV surveys, the extreme pressure and the absence of stellar radiation at the surface of the planet would render them inhospitable for life as we know it. This large atmosphere however may result from the bias to detect larger planets inherent to photometric search or simply to the large variety of planets. The bias of transit searches driven by RV observations is toward more massive rather than bigger planets. As a result, they may provide candidates with structure and composition much different from those the photometric-search finds.

\begin{acknowledgements}
We thank the anonymous referee for his careful reading and suggestions. MG is FNRS Research Associate. NCS, VN and IB acknowedge the support by the European Research Council/European Community under the FP7 through a Starting Grant, as well as in the form of grants reference PTDC/CTE-AST/098528/2008, PTDC/CTE-AST/098604/2008 and SFRH/BD/60688/2009, funded by Funda\c{c}\~ao para a Ci\^encia e a Tecnologia (FCT), Portugal. NCS would further like to thank the support from FCT through a Ci\^encia\,2007 contract funded by FCT/MCTES (Portugal) and POPH/FSE (EC).

\end{acknowledgements}

\bibliographystyle{aa}
\bibliography{mybib}

\end{document}